# CLIENT-TO-CLIENT STREAMING SCHEME FOR VOD APPLICATIONS

M Dakshayini[1], Dr T R Gopala Krishnan Nair[2]

[1]Research Scholar, Dr. MGR University.  Working Dept. of ISE, BMSCE, Bangalore. Member, Multimedia Research Group, Research Centre, DSI, Bangalore.
shantha_dakshu@yahoo.co.in

[2] Director, Research and Industry Incubation Centre, DSI, Bangalore.
trgnair@yahoo.com

## ABSTRACT

*In this paper, we propose an efficient client-to-client streaming approach to cooperatively stream the video using chaining technique with unicast communication among the clients. This approach considers two major issues of VoD 1) Prefix caching scheme to accommodate more number of videos closer to client, so that the request-service delay for the user can be minimized.  2) Cooperative proxy and client chaining scheme for streaming the videos using unicasting.  This approach minimizes the client rejection rate and bandwidth requirement on server to proxy and proxy to client path. Our simulation results show that the proposed approach achieves reduced client waiting time and optimal prefix caching of videos minimizing server to proxy path bandwidth usage by utilizing the client to client bandwidth, which is occasionally used when compared to busy server to proxy path bandwidth.*

## KEYWORDS

*Prefix Caching, Cooperative Clients, Streaming, Bandwidth Usage, Chaining.*

## 1. INTRODUCTION

Since streaming of any multimedia object like high quality video consumes a significantly large amount of network resources, network bandwidth limitation is the major constraint in most of the multimedia systems. So request-to-service delay, network traffic, congestion and server overloading are the main parameters to be considered in video streaming over the communication networks that affect the quality of service (QoS). Providing video-on-demand (VoD) service over the internet in a scalable way is a challenging problem. The difficulty is twofold: First, it is not a trivial task to stream video on an end-to-end basis because of a video's high bandwidth requirement and long duration. Second, scalability issues arise when attempting to service a large number of clients. In particular, a popular video generally attracts a large number of users that issues requests asynchronously [1].

There are many VoD schemes proposed to address these problems: patching, batching, periodical broadcasting, prefix caching and chaining.

In batching [6,8], the server makes the batches of requests for the same video together if their time of arrival  are closer, and multicasts the video to these requests to save the network I/O bandwidth. In patching [2], the main server sends the complete video clip to the first client. Later clients  join the existing multicast channel and get the missing data of the video using unicast channels. Another innovative technique is periodical broadcasting [9]. In this approach, popular videos are partitioned into a series of segments and these segments are continually broadcasted on several dedicated channels. Before clients start playing the videos, they usually have to wait for a time length equivalent to the first segment. Therefore, only near VoD service is provided.

Another promising approach to improve the bandwidth consumption issue is proxy caching [1, 4 ,5]. In this approach, there exists a proxy between a central server and client clouds. Partial video (or entire video) files are stored in proxies and the rest are stored in the central server. Proxies send cached videos to clients and request the remaining portion of the video from servers on behalf of clients.

In [3] researchers have proposed a region based hybrid algorithm for chaining, in which they reject the requests of different regions of the same proxy server, due to which rejection rate may be high. R Ashok Kumar et al.[12] have proposed a streaming scheme, in which rejection rate is very high until number of chains increases for the video. Hyunjoo and Heon in [10] have proposed another chaining scheme with VCR operations. Here they do stream the video data from main server, but they consider constant threshold value, due to which more number of clients may not be able to share the same chain, hence increasing the rejection rate.[6,8] proposes a batching technique called extended chaining (HEC), which unfairly forces the requests arriving early in a batch to wait for the latecomers. As a result, the reneging rate can be high in a system which employs this technique. To reduce the long access latency and client rejection ratio, in this paper we propose, a prefix caching based new client-to-client chaining mechanism for distributed VoD system to achieve the reduced client waiting time and the client request-rejection ratio.

This architecture for this approach consists of a centralized multimedia server [CMS] which is connected to distributed proxy servers. To each of the proxy server a cloud of users are connected.

The organization of rest of the paper is as follows: Section 2 discuses the system Model and the analysis of various parameters used in the model. In section 3 we present a proposed streaming algorithm in detail, the simulation model and performance evaluations are presented in Section 4. The section 5 presents the conclusions and the further work.

## 2. SYSTEM MODEL

The parameters considered for the model are shown in table 1. Let $N$ be a stochastic variable representing the group of videos and it may take the different values (videos) for $V_i$ ($i=1,2..N$). we assume that the client's requests arrive according to Poisson process with the arrival rate $\lambda$. Let $S_i$ be the size (duration in minutes) of $i^{th}$ video ($i=1..N$) with mean arrival rates $\lambda_1 ... \lambda_N$ respectively that are being streamed to the users using proxy server (*PS*). PS has a caching buffer large enough to cache total $B$ minutes of $K$ number of video prefixes.

Table 1. Parameters of the System Model

| Parameter | Definition |
|---|---|
| N | Total number of videos |
| $V_i$ | $i^{th}$ video (i=1..N) |
| $S_i$ | The size(minutes) of $i^{th}$ video(*i=1..N*) |
| $\lambda_i$ | mean arrival rate of $i^{th}$ video |
| PS | proxy server |
| (*Pref-1*) | $W_1$ minutes video of $V_i$ |
| B | Total size (minutes) of Proxy buffer |
| K | Total number of videos at PS |
| W | Size (Minutes) of (pref-1) |

To increase the video availability rate at LPSG, The complete video is divided into two parts. Part1; first $W_1$ minutes of each video $V_i$ is referred to as (*pref*)$_i$ of $V_i$ and is cached in any one of the proxy servers of the group only once.

Part2; remaining portion of the video $V_i$ is referred to as suffix of $V_i$.

i.e. $B = \sum_{i=1}^{K} (pref)_i$

Based on the frequency of user requests to any video, the popularity of the videos and size of *preffix* to be cached at *PS* are determined. The size (W) of *(pref))* for $i^{th}$ video is determined as follows.

$$W(pref)_i = x_i \times S_i \text{ where } 0 < x_i < 1$$

Where $x_i$ is the probability of occurrence of user requests with frequency for video *i* from last *t* minutes. This arrangement allows the PS to cache maximum portion of more number of most frequently requesting videos. Hence most of the requests can be served immediately from PS itself, which reduces the request-service delay for the user [ $(\text{Re}\,q\text{-}Ser)_{delay}$ ] and network bandwidth requirement on server-proxy path [ $BW^{S-P}$ ] significantly.

The other output stochastic variable considered is $R_{rej}$. It is the request rejection ratio, which is the ratio of the number of requests rejected ($N_{rej}$) to total number of requests arrived at the system (R), which is inversely proportional to the system throughput $S_{eff}$, where $S_{eff} = \frac{Q}{R}$ is the ratio of number of requests served (*Q*) to total number of requests arrived (*R*) at the system as shown in the figure 1. The optimization problem is to maximize $S_{eff}$ by minimizing the client rejection ratio $R_{rej}$, average Request-service delay and average bandwidth usage is.

Maximize System Efficiency $S_{eff} = \frac{Q}{R}$

Minimizing

- Avg WAN Bandwidth usage

$$BW^{S-P} = \sum_{i=1}^{Q} bw(S - (pref))_i^{S-P}$$

- Avg Request-service delay for the user

$$(\text{Re}\,q\text{-}Ser)_{delay} = \frac{1}{Q} \sum_{i=1}^{Q} (\text{Re}\,q\text{-}Ser)_{delay}$$

- Avg Request-rejection ratio

$R_{rej} = N_{rej}/R$

Subject to

$$B = \sum_{i=1}^{K} (pref - 1)_i, \quad W(pref) > 0$$

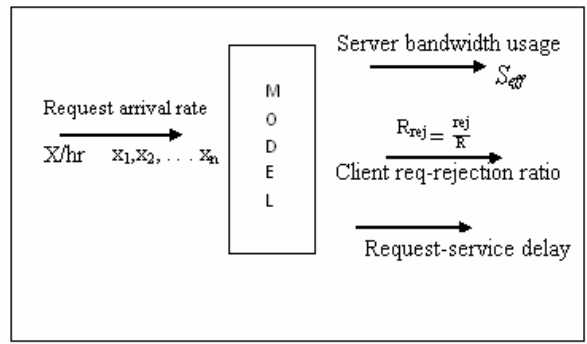

Figure 1 System Simulation Model

## 3. PROPOSED STREAMING ALGORITHM

### 3.1. Distributed VoD Algorithm

The distributed VoD architecture considered in this approach is as shown in figure 2. This architecture consists of a C*MS*, which is connected to a set of *PSs*. Each *PS* has various modules such as, *Interaction Module of PS (IM)*–Interacts with the client. *Service Manager (Ser-mgr)* - Manages the streaming of the videos. *Request handler*–Handles the requests from the user, *Client Manager (CM)* – Observes and updates the popularity of videos at PS. *SCL - streaming clients list* is the list with complete details of list of videos being streamed from that PS and the corresponding active chain of clients for that video. To each of these PSs a large number of users are connected. Each proxy is called as a parent proxy (PP) to its clients. Each client has various modules such as, *Buffer-manager*–buffers the video data received at the *Buffer*. *Media Player* –plays the buffered video. *Client Agent*– Performs the Communication with PS and other clients. LAC – *List of Applicant clients* – received from the PS. This is the list of *d* applicant clients of the active chain of the video, from whom a requested client can get the requested video stream.

The *PS* caches the prefix of videos distributed by *VDM,* and streams this cached portion of video to the client upon the request through *LAN* using its less expensive bandwidth. We assume that, proxies and their clients are closely located with relatively low communication cost [1]. The main central multimedia server in which all the videos completely stored is placed far away from PS, which involves high cost remote communication. The CMS and the *PSs* are assumed to be interconnected through high capacity optic fibre cables.

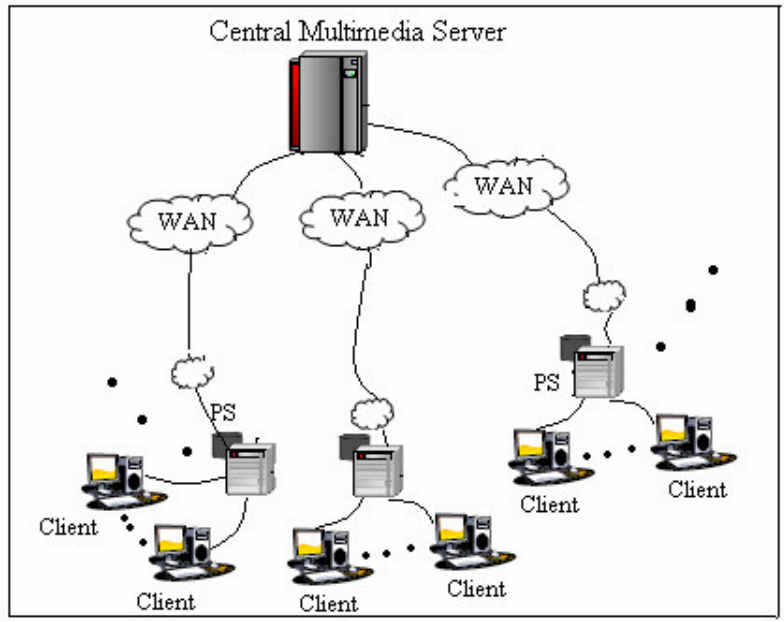

Figure 2 Distributed VoD Architecture

### 3.2. Proposed Streaming Algorithm

The main goal of our proposed streaming scheme is to make each client act as a server while it receives the video, so that the available memory and bandwidth of the clients can be utilized more efficiently. The un-scalability of traditional client-server unicast VoD service lies in the fact that the server is the only contributor and can thus become flooded by a large number of clients submissively requesting the service. In the client-server service model, the client sets up a direct connection with the server to receive the video. In this case an amount of network bandwidth requirement on server-proxy path equal to the playback rate is consumed along the route. As the number of requests increases, the bandwidth at the server and the network is heavily consumed, due to which network becomes congested and the incoming requests are eventually rejected [2]. In contrast, we propose two schemes to address these issues 1) prefix caching technique reduces the request-service delay for the user and an amount of bandwidth consumption between client and main server. 2) Streaming scheme, where the clients not only receive the requested stream, but also contribute to the overall VoD service by forwarding the stream to other clients, whose request arrives within the threshold time of *W(pref)*. In this streaming approach all the clients are treated as potential server points. When there is a request for a video $V_i$ from the client $C_k$ at a proxy server *PS*, if the requested video $V_i$ is present at PS, then the service will be provided to $C_k$ in the following stages

### 3.2.1. Client admission phase

When the request arrives at *PSq,* the R*equest-handler* (R*eq-handler*) of that proxy checks for the presence of the video at *PS_q*. If it is present then it checks the flag *IS-STREAMING* of the video $V_i$. If it is not true indicates that, there are no clients existing having streamed with the same video object. Then the R*eq-handler* informs the S*er-mgr* to provide the streaming of $V_i$ to $C_k$. the S*er-mgr* starts a new stream creating a new active chain for *Vi* and updates the streaming *clients list* (*SCL*) by adding a new entry for the video $V_i$ along with its (*pref-1*) size and making IS-STREAMING flag of *Vi* true. The format of each entry of *SCL* is <*video id - sz(pref-1)- clients list being streamed with $V_i$*>. If the flag *IS- STREAMING* is true, indicating that there are already a chain of clients being streamed with the same video. Then the R*eq-handler* looks into the corresponding entry for $V_i$ in *SCL* to check whether the inter arrival time difference of the new client and other clients of the existing active chain of $V_i$ is below the threshold T=*W(pref-1),* if so $C_k$ *i*s added to the existing chain of $V_i$ and sends the *sub list of d applicant clients* (*LAC:* list of all $C_j$ | $((t_{Cj} - t_{Ck}) \leq W(prif-1)_{Vi})$ j=1..d)  to $C_k$, along    with the *msg* to get streamed from any of the client from the list(*LAC*). *Req-handler* also sends the message (msg) to all d clients of *LAC*, so that the *client agent* of these clients, who are ready to become *parent client (PC)*  of $C_k$ sends ready signal to $C_k$ indicating that they are ready to streme the video. Then $C_k$ sends Ok-start signal to the selected client $C_j$, which is comparatively more closer *(D($C_k$,$C_j$ )*: distance between $C_k$ and $C_j$ is minimum for all *j=1..d)* and also most recently being streamed with $V_i$, then $C_k$ starts receiving the stream. This client $C_j$ becomes the *PC* of $C_k$. Fig 3 shows this scenario, where we have active chains for two videos (56 and 14). Here when the client C4 requests for the video-56, it gets the *LAC* with *C1, C2 and C3* but *C4* decides to get the video stream from *C1* as the *D(C4,C1)* is comparatively minimum. In case any problem with this client, $C_k$ can request the other clients of *LAC*. This reduces the PS involvement during playback. At the end of this client admission phase, *SCL* will have the complete details of list of videos being streamed and the corresponding chain of clients of each video. Also each client knows its applicant ancestors to change its *parent client (PC)* in case of any problem with the *PC*.
If *Vi* is not present at PS then *IM* coordinates with *CMS* and downloads the $V_i$, then is streamed to *Ck*. While streaming the *W(pref)* of $V_i$ is calculated and cached at *PS*, if sufficient space is available. Otherwise the dynamic buffer allocation algorithm [16] is executed to make room to cache this prefix of $V_i$ according to popularity.  In this case  the request-service delay may be high, but the probability of downloading the complete video from the C*MS* is very less as shown by our simulation results.This approach allows most of the requests to be served immediately from the client of the active chain of the video, which is already being served from the PS and hence reduces the *client rejection-request ratio* $R_{rej}$, *server bandwidth usage,* and *request-service delay* for the user.
Whenever the sufficient buffer and bandwidth is not available in the above operation the user request is rejected.

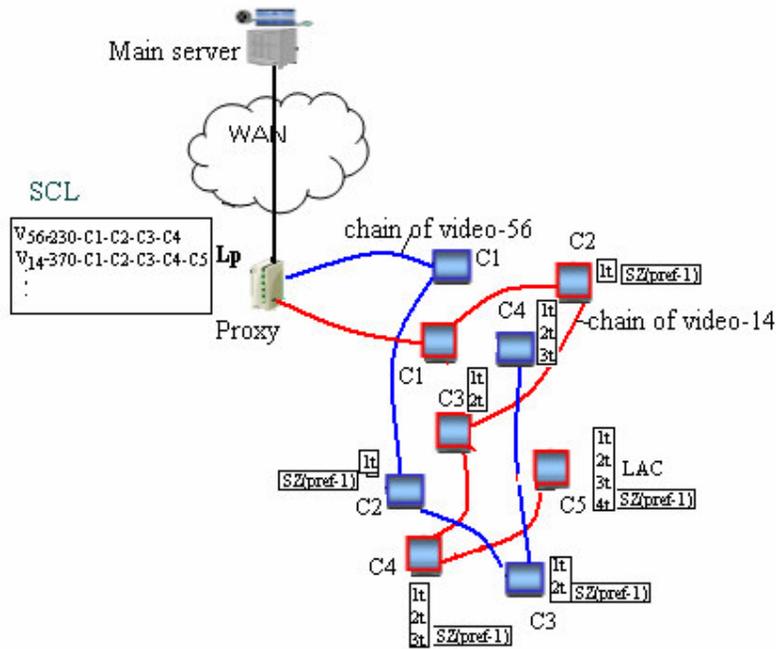

Fig. 4 chain of video-56 c1-c2-c3-c4
chain of video-14 c1-c2-c3-c4-c5

Algorithm

At Proxy server ($PS_q$)
When a request from the client $C_k$ for $V_i$ arrives at time *t* to a particular proxy *PS*, the following steps are done.
  **If** ($V_i$ € *PS*)
   Checks the *SCL*
    **If** (*IS-STREAMING*$_{Vi}$ is *TRUE*)  // $V_i$ has an Active chain
     {
       create a list of  all C*j : j=1..d*
        such that (($t_{Cj} - t_{Ck}$) ≤ (*prif*)$_{Vi}$) where *j=1..d*
       (R*eq-hdlr*)$_{PS}$  sends *msg* with *LAC* to $C_k$
     }
    **else** (*Ser-mgr s*tarts the new stream to $C_k$ and updates *SCL*)
  **else** pass the *Req*  to *CMS*
    download the $V_i$ from *CMS*
    **If** ( (*pref*)$_{Vi}$=($x_i$ ⋈ $S_i$) can be cached at *PS*)
       { cache the (*pref*)$_{Vi}$  & Start new stream to $C_k$, &
        update *SCL* }
    **else**
      {use Dynamic buffer allocation algorithm[ref.16] to make
       room to cache (*pref*)$_{Vi}$)
     **If**  (sufficient buffer is not obtained)
        Reject the request

### 3.2.2. Streaming phase

We assume that the client has sufficient buffer space and network bandwidth to accommodate the (pref) of $V_i$. Whenever the client $C_k$ requests for $V_i$, if the client is successfully admitted to the existing chain of $V_i$ then it immediately starts getting the stream from the selected client from the *LAC*, the *buffer manager (buf-mgr)* starts buffering the video clips received from its *PC*. And *media-player* starts playing it from the buffer.

Client $C_k$
$C_k$ sends the request to PS for the video $V_i$
**when** ($C_k$ receives a msg with LAC from (*req-hdlr*)$_{PS}$ )
    $C_k$ checks for the distance $D(C_k, C_j)$ with all ,$j=1..d$
    Select $C_j | D(C_k, C_j)$ is minimum &
        $((t_{Cj} - t_{Ck}) \leq W(prif)_{Vi})$
    Send Ok-Start signal to $C_j$
    Start getting streamed with the video.
**else** rejected

### 3.2.3 .Closing Phase

Any client $C_k$ in the chain may want to terminate the connection in the following two cases
Case1: Once the streaming of the entire video is finished. The *client agent* of $C_k$ has to terminate the connection with its *parent client (PC)* by sending termination signal to it, but checks whether this client is *PC* for any other clients in the chain before sending termination signal to its *parent proxy server(PPS)*. If so, it waits until the streaming of the whole video to its *child client* (*CC*) finishes or it may connect its *CC* to *PC* as explained in case2 and then sends the termination signal to its *PPS* and then terminates the connection. The *service-mgr* updates the *SCL* by removing the entry corresponding to terminated client from *SCL*.
Case2: In the middle of the streaming, if the $C_k$ wants to stop watching the video, *client-agent* of $C_k$ checks whether it is a parent for any other client in the chain. If so it checks wether $(t_{pc}(C_k) - t_{cc}(C_k))_{Vi} < (pref)_{Vi}$ is true,where $t_{pc}(C_k)$ is the arrival time of *PC* of $C_k$ and $t_{cc}(C_k)$, is the arrival time of *CC* of $C_k$. if true it sends the termination signal to its *PC*, *CC* and *PPS* indicating that it is going to be terminated and also gives the details of its *CC* to its *PC*. So the *PC* of $C_k$ can continue streaming to *CC* of $C_k$ and informs *PPS* about the change of *PC* to *CC* of $C_k$. . Then *service-mgr* of *PPS* updates the *SCL* by removing the entry for $C_k$ in active client chain list of $V_i$

## 4. SIMULATION AND RESULTS

In our simulation model we have a single *CMS* and a group six *PSs*. Each of these *PSs* is connected to 50 users. We use the client request-rejection ratio, average server bandwidth usage and the average request-service delay for the user as parameters to measure the performance of our proposed approach more efficiently.
The table 2 lists the performance parameters considered for the simulation. Here we assume that the request distribution of the videos follows a zipf-like distribution. The user request arrival rate at each *PS* is 35-50 requests per hour. The ratio of cache sizes at *CMS* and *PS* is set to $C_{CMS} : C_{PS} = 10:2$ and transmission delay between the proxy and the client as 120 to 180 ms, transmission delay between the main server and the proxy as 480 ms to 640ms, transmission delay between client to client as 60 to 90 ms, the size of the cached *(pref)* of the video as 280MB to 1120MB(25min – 1hr) in proportion to its popularity.

Table 2. Parameters of Simulation Model

| Notation | System Parameters | Default Values |
|---|---|---|
| $BW_{s-p}$ | Server-proxy bandwidth | 2 Gbps |
| $BW_{p-c}$ | Proxy-client bandwidth | 2 Mbps |
| $BW_{c-c}$ | Client-client bandwidth | 256 Kbps |
| S | Video Size | 2-3hrs, 200units/hr |
| $C_{CMS}$ | Cache Size(CMS) | 1000 |
| $C_{PS}$ | Cache Size(PS) | 200(15%) |
| $\Lambda$ | Mean request arrival rate | 44 reqs/hr |

The results of average of several simulations conducted are shown below.

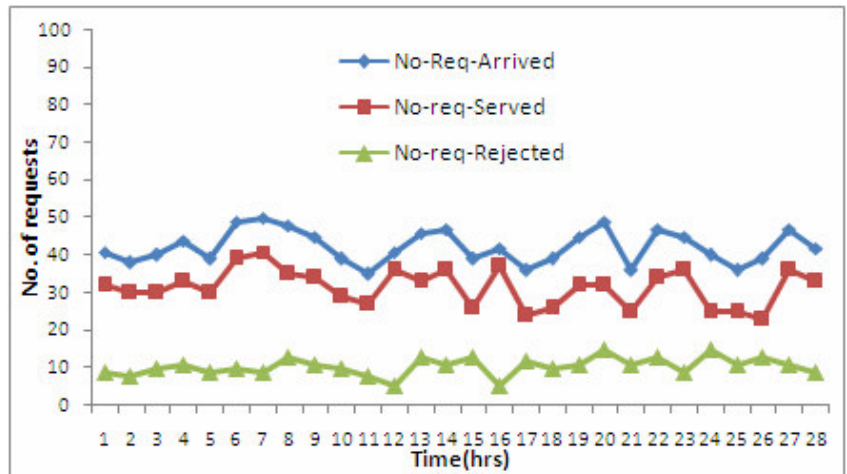

Figure 5 client rejection rate with time

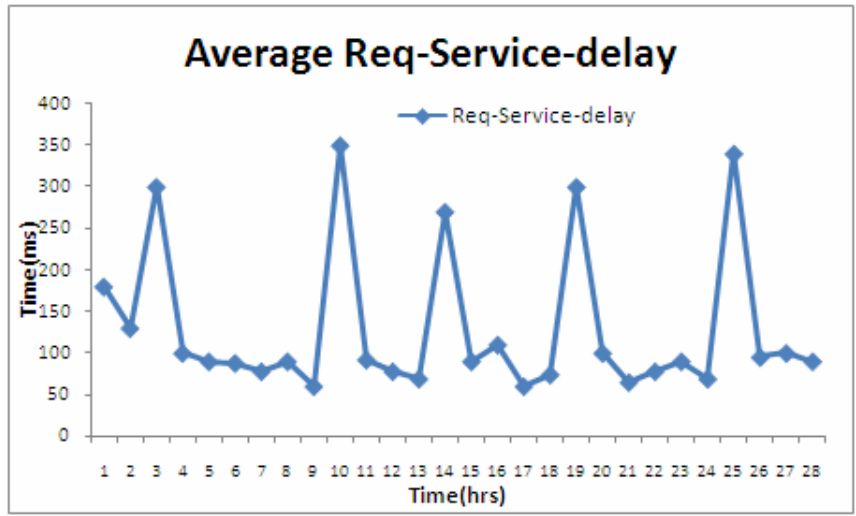

Figure 6 Request-service delay V/s Time

Figure 5 shows the client rejection ratio of our proposed system, which is very less, as our streaming scheme allows more number of clients to join the active chain immediately when their request for a video arrives, because the request arrival rate based threshold T=*W(pref)* keeps IS-STREAMING flag of the video true. This makes the requests to be served immediately from the active chain by adding them to the existing chain. This scheme significantly reduces the client rejection ratio and request-service delay for the user as shown in the Figure 6.

The graph in the figure 7 shows that, our scheme uses significantly less server bandwidth. Popularity based Prefix caching approach and dynamic buffer allocation technique with the proposed streaming scheme helps the system to cache prefix of more number of videos. Also the requests arrival rate based threshold T=*W(pref)* allows more number of clients to join the existing active chain. This maximizes bandwidth usage between the clients and hence reduces the average server bandwidth usage.

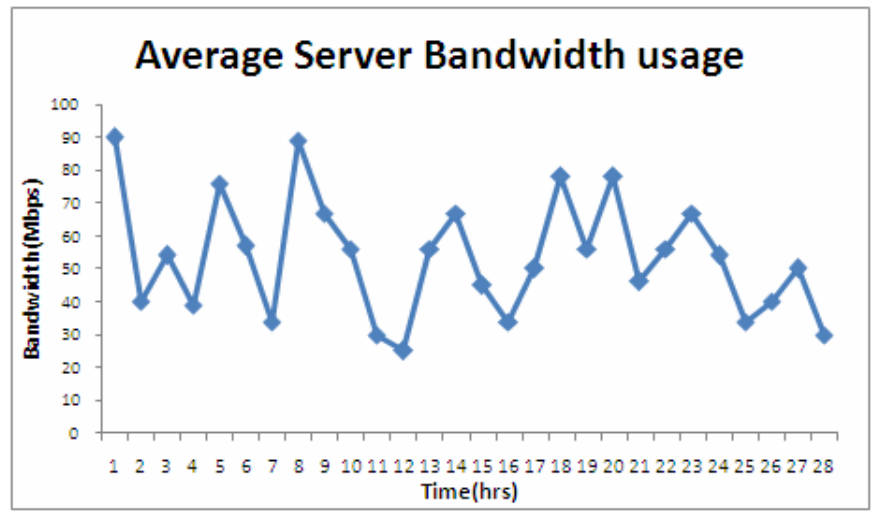

Figure 7 Bandwidth usages V/s Time

## 5. CONCLUSION

In this paper we have proposed an efficient video streaming scheme. The simulation results of this proposed scheme has demonstrated that, with the popularity based threshold W(pref-1) allows more number of clients to get benefited from the existing active chain. Also the proposed prefix caching technique with dynamic buffer allocation algorithm has significantly reduced the request rejection ratio and client waiting time. The network bandwidth usage also has been reduced by sharing the video data of the currently played video object with other clients of the active chain. The future work is being carried out to recover the failure of any client of the existing active chain for the video by introducing a network management system, which will enhance the performance our proposed scheme.

**Dr.T R Gopalakrishnan Nair** holds M.Tech. (IISc, Bangalore) and Ph.D. degree in Computer Science. He has 3 decades experience in Computer Science and Engineering through research, industry and education. He has published several papers and holds patents in multi domains. He won the PARAM Award for technology innovation. Currently he is the Director of Research and Industry in Dayananda Sagar Institutions, Bangalore, India.

**M.Dakshayini.** holds M.E in computer science. She has one and a half   decades experience in teaching field. She has published many papers. Currently she is working as a teaching faculty in the department of Information science and engineering at BMS College Of Engineering, Bangalore , India.